\documentclass[a4paper,10pt,twoside]{cpc-hepnp}
\usepackage{multicol}
\usepackage{graphicx}
\usepackage{booktabs}
\usepackage{amssymb,bm,mathrsfs,bbm,amscd}
\usepackage{lastpage}
\usepackage{changepage} 

\usepackage[scientific-notation=true]{siunitx}
\usepackage{multirow}
\usepackage{tikz}
\usetikzlibrary{shapes.geometric, arrows}

\usepackage[utf8]{inputenc}
\usepackage[colorlinks=true, linkcolor=red, citecolor=blue]{hyperref}

\usepackage{amsmath,amsfonts,amssymb,bm}
\usepackage{graphicx}
\usepackage{color}
\usepackage{subfigure}
\usepackage{multirow}
\usepackage{textcomp}
\usepackage{slashed}
\usepackage{cancel}

\usepackage{enumitem}

\def\({\left(}
\def\){\right)}
\def\[{\left[}
\def\]{\right]}

\def\be{\begin{eqnarray}}
\def\ee{\end{eqnarray}}
\usepackage[capitalise]{cleveref}
\crefname{figure}{Fig.}{Figs.}
\Crefname{figure}{Fig.}{Figs.}

\begin{document}

\fancyhead[c]{\small Submitted to Chinese Physics C} \fancyfoot[C]{\small Page-\thepage}



\title{The Decay constants of $B_c(nS)$ and $B^*_c(nS)$
}

\author{%
      Chao Sun $^{1}$, Ru-Hui Ni $^{1}$,
      Muyang Chen $^{1,2,\dag}$ \email{$\dag$ muyang@hunnu.edu.cn}
\footnote{This work is supported by the National Natural Science Foundation of China under contracts No. 12005060.}
}

\maketitle

\address{%
$^1$ Department of Physics, Hunan Normal University, and Key Laboratory of Low-Dimensional
Quantum Structures and Quantum Control of Ministry of Education, Changsha 410081, China\\
$^2$ Synergetic Innovation Center for Quantum Effects and Applications (SICQEA), Hunan Normal University, Changsha 410081, China
}

\begin{abstract}
The decay constants of the low lying S-wave $B_c$ mesons, i.e. $B_c(nS)$ and $B^*_c(nS)$ with $n\leq 3$, are calculated in the nonrelativistic quark model. The running coupling of the strong interaction is taken into account, and the uncertainty due to varying parameters and losing Lorentz covariance are considered carefully. As a byproduct, the decay constants of the low lying S-wave charmonium and bottomium states are given in the appendixes.
\end{abstract}




\begin{multicols}{2}

\section{Introduction}
\label{sec:intro} 
\noindent The $B_c$ mesons are the only open flavor mesons containing two heavy valence quarks, i.e. one charm quark and one bottom anti-quark (or vice versa). The flavor forbids their annihilation into gluons or photons, so that the ground state pseudoscalar $B_c(1S)$ can only decay weakly, which makes it particularly interesting for the study of weak interaction. In the experimental aspect, the $B_c$ mesons are much less explored than the charmonium and bottomonium due to the small production rate, as the
dominant production mechanism requires the production of both $c\overline{c}$ and $b\overline{b}$ pairs. The $B_c(1S)$ meson was first observed by CDF experiment in 1998 \cite{Abe1998}. In the later years, the mass and life time of $B_c(1S)$ were measured precisely, and its hadronic decay modes were also observed \cite{Abulencia2006,Aaltonen2008,Abazov2008,Abazov2009}. The excited $B_c$ meson state was not observed until 2014, by the ATLAS experiment \cite{Aad2014}. The mass of $B_c(2S)$ was measured by LHCb experiment \cite{Aaij2019} and CMS experiment \cite{Sirunyan2019} independently in 2019. However, for the vector $B_c$ mesons, only the mass difference $M_{B_c^*(2S)} - M_{B_c^*(1S)} = 567\text{ MeV}$ is known \cite{Sirunyan2019}.

In the theoretical aspect, the mass spectrum and the decays of $B_c$ mesons are investigated by various methods. For example, the quark model \cite{Ortega2020,Li2019,Hernandez2006,Godfrey2004,Ebert2003,Chang1998}, the light-front quark model \cite{Shi2016,Wang2009,Choi2009}, the QCD sum rule \cite{Aliev2019,Kiselev2000}, the QCD factorization \cite{Feng2022,Qiao2014,Liu2010,Choi2009,Chang2004}, the instantaneous approximation Bethe-Salpeter equation \cite{Wang2022,Ding2021}, the continuum QCD approach \cite{Chen2021,Chen2020,Chen2019}, the lattice QCD \cite{Mathur2018} and other methods \cite{Chang2004b,Chang2004a,Chang2001}. The quark model, with the interaction motivated by quantum chromodynamics (QCD), is quite successful in describing the hadron spectrum and decay branching ratios, see Refs. \cite{Capstick1985,Godfrey1985} for an introduction. The nonrelativistic version of the quark model is suitable for heavy quark systems. It is not only phenomenological successful in describing mesons and baryons \cite{Deng2017,Deng2017a,Liu2020}, but also powerful in predicting the properties of exotic hadrons, such as tetraquarks \cite{Liu2019,Liu2021}.

The decay constant carries the information of strong interaction in leptonic decay, and thus it is intrinsically nonperturbative. A presice determination of the decay constant is crucial for a precise calculation of the leptonic decay width. In this paper, we investigate the decay constants of low lying S-wave $B_c$ mesons, i.e. $B_c(nS)$ and $B^*_c(nS)$ with $n\leq 3$ in the nonrelativistic quark model. As the $B_c$ mesons are less explored, our result is significant for both theoretical and experimental exploring of the $B_c$ family. The work of Lakhina and Swanson \cite{Lakhina2006} showed that two elements are important in calculating decay constants within nonrelativistic quark model, one is the running coupling of strong interaction, the other is the relativistic correction. Both of these two elements are taken into account in this paper. What's more, the uncertainty due to varying parameters and losing Lorentz covariance are considered carefully.

This paper is organized as following. In section \ref{sec:model}, we introduce the framework of the quark model. The formulas for the decay constants in quark model are given in section \ref{sec:decayconstant}. In section \ref{sec:results}, the results of mass spectrum and decay constants are presented and discussed. Summary and conclusions are given in section \ref{sec:conclusion}. We also present the mass spectrum and decay constants of charmonium in Appendix A and those of bottomium in \ref{sec:appendixB} Appendix B for comparison.

\section{The model}\label{sec:model}
\noindent The framework has been introduced elsewhere, see for example Refs. \cite{Deng2017,Deng2017a,Li2019}. We recapitulate the framework here for completeness and to specify the details. The masses and wave functions are obtained by solving the radial Schrödinger equation, 
\begin{equation}\label{eq:SchrodingerEq}
 \frac{1}{r^2}\frac{\text{d}}{\text{d}r}(r^2\frac{\text{d}}{\text{d}r})R(r) + \left[ 2\mu_m (E-V) - \frac{L(L+1)}{r^2} \right]R(r)= 0,
\end{equation}
where $r$ is the distance between the two constituent quarks, $R(r)$ is the radial wave function, $\mu_m = \frac{m \bar{m}}{m + \bar{m}}$ is the reduced mass with $m$ and $\bar{m}$ being the constituent quark masses, $L$ is the orbital angular moment quantum number. $V$ is the potential between the quarks and $E$ is the energy of this system. Then the meson mass is $M = m + \bar{m} + E$. Note that the complete wave function is $\Phi_{nLM_L}(\bm{r}) = R_{nL}(r)Y_{LM_L}(\theta,\phi)$, where $n$ is the main quantum number, $M_L$ is the magnetic quantum number of orbital angular momentum, and $ Y_{LM_L}(\theta,\phi)$ is the spheric harmoics. In this paper a bold character stands for a 3-dimension vector, for example, $\bm{r} = \vec{r}$.

The potential could be decomposed into
\begin{equation}\label{eq:interaction}
 V = H^{\text{SI}} + H^{\text{SS}} + H^{\text{T}} + H^{\text{SO}}.
\end{equation}
$H^{\text{SI}}$ is the spin independent part, which is composed of a coulombic potential and a linear potential,
\begin{equation}\label{eq:interactionSI}
 H^{\text{SI}} = -\frac{4\alpha_s(Q^2)}{3r} + br,
\end{equation}
where $b$ is a constant and $\alpha_s(Q^2)$ is the running coupling of the strong interaction. The other three terms are spin dependent.
\begin{equation}\label{eq:interactionSS}
 H^{\text{SS}} = \frac{32\pi\alpha_s(Q^2)}{9m\bar{m}}\tilde{\delta}_\sigma(\bm{r}) \bm{s}\cdot\bm{\bar{s}}
\end{equation}
is the spin-spin contact hyperfine potential, where $\bm{s}$ and $\bm{\bar{s}}$ are the spin of the quark and antiquark respectively, and $\tilde{\delta}_\sigma(\bm{r}) = (\frac{\sigma}{\sqrt{\pi}})^3 \text{e}^{-\sigma^2 r^2}$ with $\sigma$ being a parameter.
\begin{equation}\label{eq:interactionT}
 H^{\text{T}} = \frac{4\alpha_s(Q^2)}{3m\bar{m}} \frac{1}{r^3}\left( 3\frac{(\bm{s}\cdot\bm{r})(\bm{\bar{s}}\cdot\bm{r})}{r^2} - \bm{s}\cdot\bm{\bar{s}} \right)
\end{equation}
is the tensor potential. $H^{\text{SO}}$ is the spin-orbital interaction potential and could be decomposed into a symmetric part $H^{\text{SO+}}$ and an anti-symmetric part $H^{\text{SO-}}$, i.e.
\begin{eqnarray}\label{eq:interactionSO}
 H^{\text{SO}} &=& H^{\text{SO+}} + H^{\text{SO-}},\\\nonumber
 H^{\text{SO+}}  &=&  \frac{\bm{S}_+\cdot\bm{L}}{2}\left[ (\frac{1}{2m^2} + \frac{1}{2\bar{m}^2})(\frac{4\alpha_s(Q^2)}{3r^3} - \frac{b}{r})\right. \\\label{eq:interactionSO+}
 && \left. + \frac{8\alpha_s(Q^2)}{3m\bar{m}r^3} \right],\\\label{eq:interactionSO-}
 H^{\text{SO-}}  &=&  \frac{\bm{S}_-\cdot\bm{L}}{2}\left[ (\frac{1}{2m^2} - \frac{1}{2\bar{m}^2})(\frac{4\alpha_s(Q^2)}{3r^3} - \frac{b}{r}) \right],
\end{eqnarray}
where $\bm{S}_\pm = \bm{s} \pm \bm{\bar{s}}$, and $\bm{L}$ is the orbital angular momentum of the quark and antiquark system.

In equations (\ref{eq:interactionSI})$\sim$(\ref{eq:interactionSO-}), the running coupling takes the following form,
\begin{equation}\label{eq:runningAlphas}
 \alpha_s(Q^2) = \frac{4\pi}{\beta \log(\text{e}^{\frac{4\pi}{\beta\alpha_0}} + \frac{Q^2}{\Lambda^2_{\text{QCD}}})},
\end{equation}
where $\Lambda_{\text{QCD}}$ is the energy scale below which nonperturbative effects take over, $\beta = 11-\frac{2}{3}N_f$ with $N_f$ being the flavor number, $Q$ is the typical momentum of the system, and $\alpha_0$ is a constant. Eq. (\ref{eq:runningAlphas}) approaches the one loop running form of QCD at large $Q^2$ and saturates at low $Q^2$. In practice $\alpha_s(Q^2)$ is parametrized by the form of a sum of Gaussian functions and transformed into $\alpha_s(r)$ as in Ref. \cite{Godfrey1985}.

It should be mentioned that the potential containing $\frac{1}{r^3}$ is divergent. Following Refs. \cite{Deng2017,Deng2017a}, a cutoff $r_c$ is introduced, so that $\frac{1}{r^3} \to \frac{1}{r^3_c}$ for $r \leq r_c$. Herein $r_c$ is a parameter to be fixed by observables. Most of the interaction operators in Eq. (\ref{eq:interaction}) are diagonal in the space with basis $|JM_J;LS \rangle$ except $H^{\text{SO-}}$ and $H^{\text{T}}$, where $J$, $L$ and $S$ are the total, orbital and spin angular momenta quantum number, $M_J$ is the magnetic quantum number. The anti-symmetric part of the spin-orbital interaction, $H^{\text{SO-}}$, arising only when the quark masses are unequal, causes $^3L_J \leftrightarrow ^1L_J$ mixing. The tensor interaction, $H^{\text{T}}$, causes $^3L_J \leftrightarrow ^3L'_J$ mixing. The former mixing is considered in our calculation while the later one is ignored, as the mixing due to the tensor interaction is very weak \cite{Godfrey1985}. 

There are 8 parameters in all: $m$, $\bar{m}$, $N_f$, $\Lambda_{\text{QCD}}$, $\alpha_0$, $b$, $\sigma$ and $r_c$. $m$ and $\bar{m}$ are fixed by the mass spectrum of charmonium and bottomium, see Appendix A and Appendix B. $N_f$ and $\Lambda_{\text{QCD}}$ are chosen according to QCD estimation. $N_f=4$ for charmonium and $B_c$ mesons, and $N_f = 5$ for bottomium mesons. In this work we vary $\Lambda_{\text{QCD}}$ in the range $0.2 \text{ GeV} <\Lambda_{\text{QCD}} < 0.4\text{ GeV}$, then $\alpha_0$, $b$, $\sigma$ and $r_c$ are fixed by the masses of $B_c(1^1S_0)$, $B_c(2^1S_0)$, $B^*_c(1^3S_1)$ and $B_c(1^3P_0)$. For the $B_c$ meson masses, the experiment values \cite{Zyla2020} or the lattice QCD results \cite{Mathur2018} are referred, see Table \ref{tab:mcb}.

\section{Decay constant}\label{sec:decayconstant}
\noindent The decay constant of a pseudoscalar meson, $f_P$, is defined by
\begin{equation}\label{eq:decayfp}
 p^\mu f_P \text{e}^{-ip\cdot x} = i\langle 0| j^{\mu 5}(x) |P(p) \rangle,
\end{equation}
where $|P(p) \rangle$ is the pseudoscalar meson state, $p^\mu$ is the meson 4-momentum, and $j^{\mu 5}(x) = \bar{\psi}\gamma^\mu \gamma^5 \psi(x)$ is the axial vector current with $\psi(x)$ being the quark field. In quark model the pseudoscalar meson state is described by
\begin{eqnarray}\nonumber
 |P(p) \rangle &=& \sqrt{\frac{2E_p}{N_c}}\chi^{\bm{SM_S}}_{\bm{s\bar{s}}} \int\frac{d^3\bm{k} d^3\bm{\bar{k}}}{(2\pi)^3}\Phi\left(\frac{\bar{m}\bm{k} - m\bm{\bar{k}}}{m+\bar{m}}\right) \\\label{eq:state}
 && \cdot\delta^{(3)}(\bm{k}+\bm{\bar{k}}-\bm{p}) b^\dag_{\bm{ks}} d^\dag_{\bm{\bar{k}\bar{s}}}| 0 \rangle,
 \end{eqnarray}
where $\bm{k}$, $\bm{\bar{k}}$ and $\bm{p}$ are the momentum of the quark, antiquark and meson respectively, $E_p =\sqrt{M^2 + \bm{p}^2}$ is the meson energy, $N_c$ is the color number, $\bm{S}(=\bm{S}_+)$ is the total spin and $\bm{M_S}$ is its z-projection (in the case of pseudoscalar meson, $\bm{S} = \bm{M_S} = 0$), $b^\dag_{\bm{ks}}$ and $d^\dag_{\bm{\bar{k}\bar{s}}}$ are the creation operator of the quark and antiquark respectively. $\chi^{\bm{SM_S}}_{\bm{s\bar{s}}}$ is the spin wave function, and $\Phi\left(\frac{\bar{m}\bm{k} - m\bm{\bar{k}}}{m+\bar{m}} = \bm{k}_r\right)$ is wave function in momentum space, $\bm{k}_r$ is the relative momentum between the quark and antiquark. While $\Phi(\bm{k}_r) = \int \text{d}^3\bm{r}\Phi(\bm{r}) \text{e}^{-i\bm{k}_r \cdot\bm{r}}$, we use the same symbol for wave functions in coordinate space and momentum space.

The decay constant is Lorentz invariant by definition, Eq. (\ref{eq:decayfp}). However, $|P(p) \rangle$ defined by Eq. (\ref{eq:state}) is not Lorentz covariant, and thus leads to ambiguity about the decay constant. Let the 4-momentum to be $p^\mu = (E_p, \bm{p})$ and $\bm{p} = (0,0,p)$, we could obtain the decay constant by comparing the temporal ($\mu = 0$) component or the spacial ($\mu = 3$) component of Eq. (\ref{eq:decayfp}). The decay constant obtained with temporal component is
\begin{eqnarray}\nonumber
  f_P &=& \sqrt{\frac{N_c}{E_p}} \int \frac{\text{d}^3\bm{l}}{(2\pi)^3} \Phi(\bm{l})\sqrt{(1+\frac{m}{E_{l_+}}) (1+\frac{\bar{m}}{\bar{E}_{l_-}})} \\\label{eq:fptemporal}
&&\times\left[ 1-\frac{\bm{l}_+\cdot \bm{l}_-}{(E_{l_+}+m) (\bar{E}_{l_-}+\bar{m})} \right],
\end{eqnarray}
where $\bm{l}_+ = \bm{l} + \frac{m\bm{p}}{m+\bar{m}}$, $\bm{l}_- = \bm{l} - \frac{\bar{m}\bm{p}}{m+\bar{m}}$, $E_{l_+} = \sqrt{(\bm{l}_+)^2 + m^2}$, and $\bar{E}_{l_-} = \sqrt{(\bm{l}_-)^2 + \bar{m}^2}$.
The decay constant obtained with spacial component is
\begin{eqnarray}\nonumber
  f_P &=& \frac{\sqrt{N_c E_p}}{\bm{p}^2} \int \frac{\text{d}^3 \bm{l}}{(2\pi)^3} \Phi(\bm{l}) \sqrt{(1+\frac{m}{E_{l_+}}) (1+\frac{\bar{m}}{\bar{E}_{l_-}})}\\\label{eq:fpspacial}
 &&\times \left[ \frac{\bm{p}\cdot\bm{l}_+}{E_{l_+}+m} - \frac{\bm{p}\cdot\bm{l}_-}{ \bar{E}_{l_-}+\bar{m} } \right].
\end{eqnarray}
The Lorentz covariance is violated in two aspects. Firstly, Eqs. (\ref{eq:fptemporal}) and (\ref{eq:fpspacial}) lead to different results. Secondly, $f_P$ varies as the momentum $p = |\bm{p}|$ varies. Losing Lorentz covariance is a deficiency of nonrelativistic quark model and covariance is only recovered in the nonrelativistic and weak coupling limits \cite{Lakhina2006}. Herein we treat the center value as the prediction, and the deviation is treated as the uncertainty due to losing Lorentz covariance.

The decay constant of a vector meson, $f_V$, is defined by
\begin{equation}\label{eq:decayfv}
 M_V f_V \epsilon^\mu \text{e}^{-ip\cdot x} = \langle 0| j^{\mu}(x) |V(p) \rangle,
\end{equation}
where $M_V$ is the vector meson mass, $\epsilon^\mu$ is its polarization vector, $j^{\mu}(x) = \bar{\psi}\gamma^\mu \psi(x)$ is the vector current, the vector meson state is the same as Eq. (\ref{eq:state}) except $\bm{S} = 1$ and $\bm{M_S} = 0, \pm 1$ (we use the quantum number to present the value of the angular momentum). With $p^\mu = (E_p, 0,0,p)$, the polarization vector is
\begin{eqnarray}
 \epsilon^\mu_+ = (0, -\frac{1}{\sqrt{2}}, -\frac{i}{\sqrt{2}}, 0) &\text{for}& \bm{M_S} = + 1,\\
 \epsilon^\mu_0 = (\frac{p}{M_V}, 0, 0, \frac{E_p}{M_V}) &\text{for}& \bm{M_S} = 0,\\
 \epsilon^\mu_- = (0, \frac{1}{\sqrt{2}}, -\frac{i}{\sqrt{2}}, 0) &\text{for}& \bm{M_S} = - 1.
 \end{eqnarray}
We will get three different expressions for $f_V$ in nonrelativistic quark model. Let $\epsilon^\mu = \epsilon^\mu_0$ and $\mu=0$ (temporal), 
\begin{eqnarray}\nonumber
 f_V &=& \frac{\sqrt{N_c E_p}}{\bm{p}^2} \int \frac{\text{d}^3 \bm{l}}{(2\pi)^3} \Phi(\bm{l}) \sqrt{(1+\frac{m}{E_{l_+}}) (1+\frac{\bar{m}}{\bar{E}_{l_-}})}\\\label{eq:fvtemporal}
 && \times \left[ \frac{\bm{p}\cdot\bm{l}_+}{E_{l_+}+m} - \frac{\bm{p}\cdot\bm{l}_-}{ \bar{E}_{l_-}+\bar{m} } \right].
\end{eqnarray}
Let $\epsilon^\mu = \epsilon^\mu_0$ and $\mu=3$ (spacial longitudinal), 
\begin{eqnarray}\nonumber
 f_V &=& \sqrt{\frac{N_c}{E_p}} \int \frac{\text{d}^3 \bm{l}}{(2\pi)^3} \Phi(\bm{l}) \sqrt{(1+\frac{m}{E_{l_+}}) (1+\frac{\bar{m}}{\bar{E}_{l_-}})} \\\label{eq:fvspacialL}
&& \times\left[ 1+\frac{ 2\bm{l}^2 -\bm{l}_+\cdot \bm{l}_- -2(\bm{l}\cdot \bm{p})^2/\bm{p}^2 }{(E_{l_+}+m) (\bar{E}_{l_-}+\bar{m})} \right].
\end{eqnarray}
Let $\epsilon^\mu = \epsilon^\mu_+ \text{ or } \epsilon^\mu_-$ and $\mu=1 \text{ or } 2$  (spacial transverse), 
\begin{eqnarray}\nonumber
 f_V &=& \frac{\sqrt{N_c E_p}}{M_V} \int \frac{\text{d}^3 \bm{l}}{(2\pi)^3} \Phi(\bm{l}) \sqrt{(1+\frac{m}{E_{l_+}}) (1+\frac{\bar{m}}{\bar{E}_{l_-}})} \\\label{eq:fvspacialT}
&& \times\left[ 1+\frac{ -\bm{l}^2 +\bm{l}_+\cdot \bm{l}_- +(\bm{l}\cdot \bm{p})^2/\bm{p}^2 }{(E_{l_+}+m) (\bar{E}_{l_-}+\bar{m})} \right].
\end{eqnarray}
Again the center value is treated as the prediction of $f_V$, and the deviation is treated as the uncertainty due to losing Lorentz covariance.

\section{Results}\label{sec:results}
\noindent We take Eq. (\ref{eq:SchrodingerEq}) as an eigenvalue problem, and solve it using the gaussian expansion method \cite{Hiyama2003}. Three parameter sets are used in our calculation, which are listed in Table \ref{tab:parameters}. The $B_c$ mass spectra corresponding to these three parameter sets are listed in Table \ref{tab:mcb} from column three to five. The parameters are fixed by the masses of $B_c(1^1S_0)$, $B_c(2^1S_0)$, $B^*_c(1^3S_1)$ and $B_c(1^3P_0)$, where the experiment values \cite{Zyla2020} (column seven) or the lattice QCD results \cite{Mathur2018} (column eight) are referred. The others are all outputs of the quark model explained from Eqs. (\ref{eq:interaction}) to (\ref{eq:runningAlphas}). We also list the result of a previous nonrelativistic quark model \cite{Li2019} using a constant $\alpha_s$ in column six.
Comparing the results using different parameters, we see that the deviation is larger as $n$ is larger. The deviation from the center value is about $30\text{ MeV}$ for $3S$ states and $50\text{ MeV}$ for $3P$ states.

\end{multicols}

\begin{center}
\tabcaption{\label{tab:parameters} Three paremater sets used in our calculation. $m_c$ and $m_b$ are fixed by the mass spectrum of charmonium and bottomium respectively, see Table \ref{tab:mcc} and Table \ref{tab:mbb} in the appendix. $N_f$ and $\Lambda_{\text{QCD}}$ are chosen according to QCD estimation. $\alpha_0$, $b$, $\sigma$ and $r_c$ are fixed by the masses of $B_c(1^1S_0)$, $B_c(2^1S_0)$, $B^*_c(1^3S_1)$ and $B_c(1^3P_0)$ (the experiment values \cite{Zyla2020} or the lattice QCD results \cite{Mathur2018} are referred).}
\centering

\renewcommand\arraystretch{1.2}
\begin{tabular}{p{5em}<{\centering}|p{3em}<{\centering}|p{3em}<{\centering}|p{3em}<{\centering}|p{3em}<{\centering}|p{3em}<{\centering}|p{3em}<{\centering}|p{3em}<{\centering}|p{3em}<{\centering}}
\hline    
  & $m_c$ (GeV)& $m_b$ (GeV)& $N_f$ & $\Lambda_{\text{QCD}}$ (GeV)& $\alpha_0$& $b$  (GeV$^2$)& $\sigma$ (GeV) & $r_c$ (fm)\\
\hline
Parameter1 & 1.591 & 4.997 & 4 & 0.20 &1.850 & 0.1515 & 1.86 & 0.538\\
Parameter2 & 1.591 & 4.997 & 4 & 0.30 &1.074 & 0.1250 & 1.50 & 0.420\\
Parameter3 & 1.591 & 4.997 & 4 & 0.40 &0.865 & 0.1126 & 1.40 & 0.345\\
\hline
\end{tabular}

\end{center}

\begin{center}
 \tabcaption{\label{tab:mcb} Mass spectrum of $B_c$ mesons (in GeV). The third to fifth columns are our results corresponding to the three parameter sets in Table \ref{tab:parameters}, where the underlined values are used to fix $\alpha_0$, $b$, $\sigma$ and $r_c$. The sixth column is the result of a previous nonrelativistic quark model using a constant $\alpha_s$. $M^{\textmd{expt.}}_{c\bar{b}}$ is the experiment value, $M_{B_c(1^1S_0)}$ and $M_{B_c(2^1S_0)}$ are taken from Ref. \cite{Zyla2020}, $M_{B^*_c(2^3S_1)}$ is obtained by combining the experiment value $M_{B^*_c(2^3S_1)} - M_{B^*_c(1^3S_1)} = 0.567 \text{ GeV}$ \cite{Sirunyan2019} and the lQCD value of $M_{B^*_c(1^3S_1)}$. $M^{\textmd{lQCD}}_{c\bar{b}}$ is the recent lattice QCD result \cite{Mathur2018}.}
\begin{tabular}{c|c|c|c|c|c|c|c}
\hline    
\multirow{2}{*}{state}&\multirow{2}{*}{$J^{\textmd{P}}$} &	\multicolumn{3}{c|}{$M_{c\bar{b}}$}	& \multirow{2}{*}{$M_{c\bar{b}}$ \textsuperscript{\cite{Li2019}}} & \multirow{2}{*}{$M^{\textmd{expt.}}_{c\bar{b}}$ \textsuperscript{\cite{Sirunyan2019,Zyla2020}}} &\multirow{2}{*}{$M^{\textmd{lQCD}}_{c\bar{b}}$ \textsuperscript{\cite{Mathur2018}}}\\
\cline{3-5}
&&  Parameter1&  Parameter2	&  Parameter3 && \\
\hline
$B_c(1^1S_0)$& $0^{-}$ & \underline{6.275} & \underline{6.275}& \underline{6.275}& 6.271 &  6.274(0.3) & 6276(3)(6) \\
$B_c(2^1S_0)$& $0^{-}$ & \underline{6.872} & \underline{6.872}& \underline{6.872}& 6.871&  6.871(1) & --\\
$B_c(3^1S_0)$& $0^{-}$ & 7.272 & 7.241& 7.220& 7.239 &  -- & --\\
$B^*_c(1^3S_1)$& $1^{-}$ & \underline{6.333} & \underline{6.333}& \underline{6.333}& 6.326&  -- & 6331(4)(6)\\
$B^*_c(2^3S_1)$& $1^{-}$ & 6.900 & 6.895& 6.893& 6.890 &  6.898(6) & --\\
$B^*_c(3^3S_1)$& $1^{-}$ & 7.292 & 7.256& 7.233& 7.252 &  -- & --\\
$B_c(1^3P_0)$& $0^{+}$ & \underline{6.712} & \underline{6.712}& \underline{6.712}& 6.714 &  -- & 6712(18)(7)\\
$B_c(2^3P_0)$& $0^{+}$ & 7.145 & 7.123& 7.106& 7.107 &  -- & --\\
$B_c(3^3P_0)$& $0^{+}$ & 7.487 & 7.433& 7.396& 7.420 &  -- & --\\
$B_c(1P_1)$& $1^{+}$ & 6.729 & 6.736& 6.744& 6.757 &  -- & 6736(17)(7)\\
$B_c(1P'_1)$& $1^{+}$ & 6.725 & 6.741& 6.755& 6.776 &  -- & --\\
$B_c(2P_1)$& $1^{+}$ & 7.153 & 7.134& 7.123& 7.134 &  -- & --\\
$B_c(2P'_1)$& $1^{+}$ & 7.145 & 7.130& 7.120& 7.150 &  -- & --\\
$B_c(3P_1)$& $1^{+}$ & 7.493 & 7.440& 7.406& 7.441 &  -- & --\\
$B_c(3P'_1)$& $1^{+}$ & 7.485 & 7.435& 7.404& 7.458 &  -- & --\\
$B_c(1^3P_2)$& $2^{+}$ & 6.735 & 6.755& 6.772& 6.787 &  -- & --\\
$B_c(2^3P_2)$& $2^{+}$ & 7.152 & 7.139& 7.133& 7.160 &  -- & --\\
$B_c(3^3P_2)$& $2^{+}$ & 7.491 & 7.441& 7.413& 7.464 &  -- & --\\
\hline
\end{tabular}

\end{center}

\begin{center}
 \tabcaption{\label{tab:mixingAngle} Mixing angles of the $nP_1$ and  $nP'_1$ (n = 1, 2, 3) states. The second to fourth columns are our results corresponding to the three parameter sets in Table \ref{tab:parameters}. The fifth column is the result of a previous quark model using a constant strong coupling \cite{Li2019}.}
\begin{tabular}{c|c|c|c|c}
\hline    
\multirow{2}{*}{Mixing Angle}& \multicolumn{3}{c|}{herein}	& \multirow{2}{*}{previous \textsuperscript{\cite{Li2019}}}\\
\cline{2-4}
&  Parameter1&  Parameter2	&  Parameter3 & \\
\hline
$\theta_{1P}$ & $30.8^\circ$& $37.3^\circ$& $34.0^\circ$& $35.5^\circ$ \\
$\theta_{2P}$ & $24.2^\circ$& $9.9^\circ$& $29.9^\circ$& $38.0^\circ$ \\
$\theta_{3P}$ & $22.0^\circ$& $14.1^\circ$& $3.6^\circ$& $39.7^\circ$ \\
\hline
\end{tabular}

\end{center}

\begin{center} 
 \includegraphics[width=0.49\textwidth]{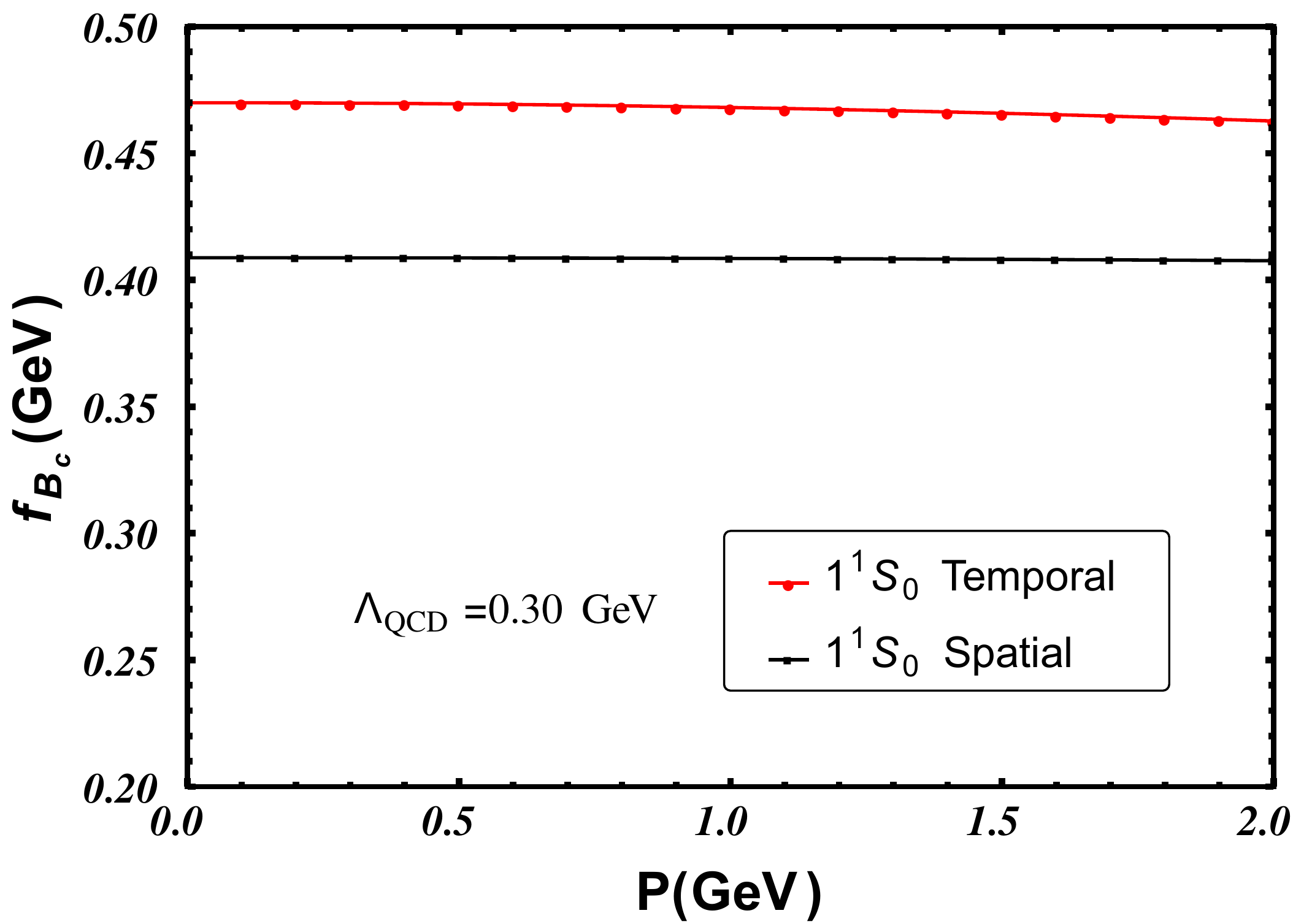}
 \includegraphics[width=0.49\textwidth]{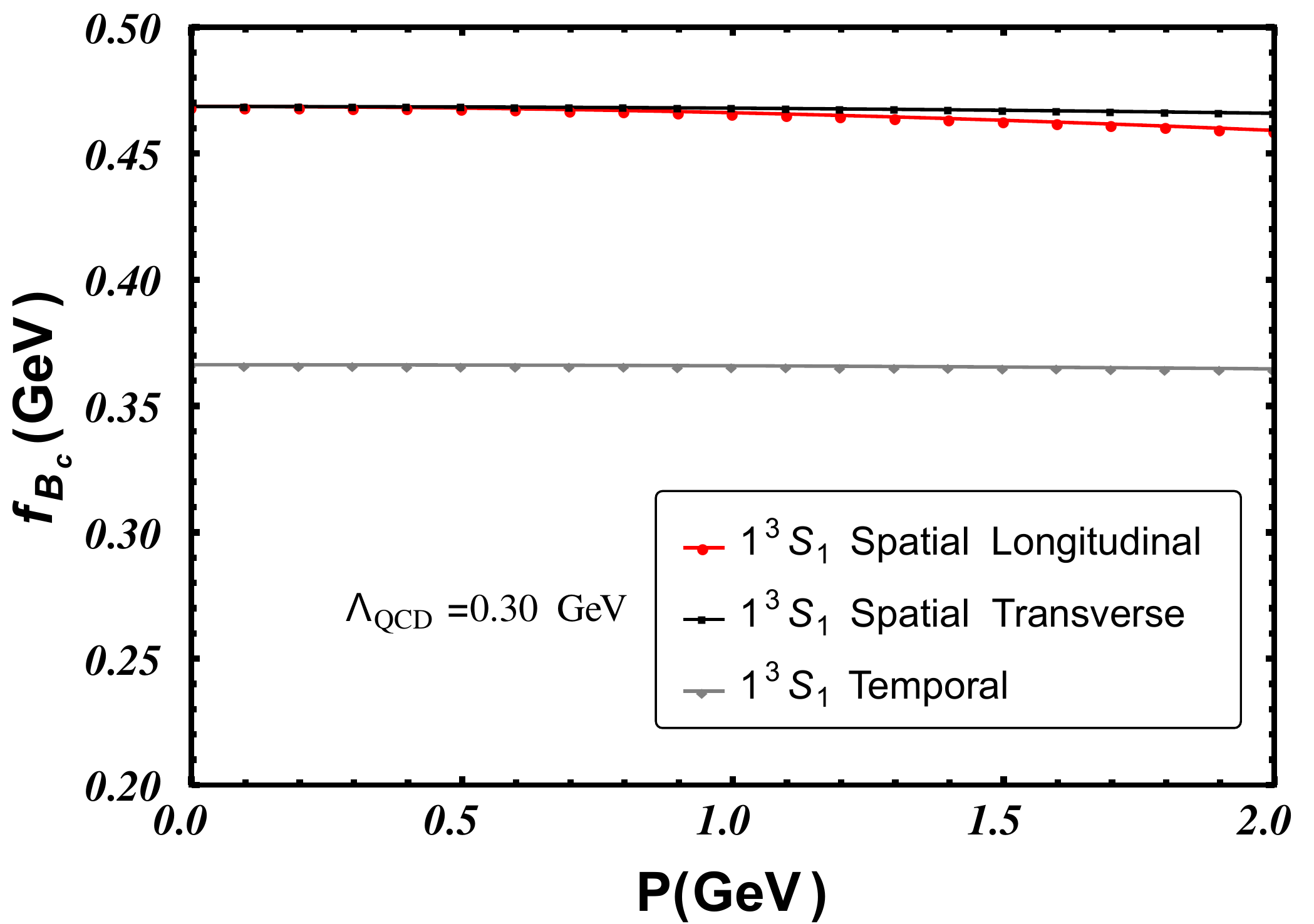}
\figcaption{\label{fig:fp} (Color Online) Decay constants calculated using Parameter2 in Table \ref{tab:parameters}, the horizontal ordinate is the moment of the meson. Left: decay constant of $B_c(1^1S_0)$, ``$1^1S_0$ Temporal" is calculated from Eq. (\ref{eq:fptemporal}), ``$1^1S_0$ Spacial" is calculated from Eq. (\ref{eq:fpspacial}). Right: decay constant of $B_c(1^3S_1)$, ``$1^3S_1$ Temporal" is calculated from Eq. (\ref{eq:fvtemporal}), ``$1^3S_1$ Spacial Longitudinal" is calculated from Eq. (\ref{eq:fvspacialL}), ``$1^3S_1$ Spacial Transverse" is calculated from Eq. (\ref{eq:fvspacialT}).}
\end{center}

\begin{center}
\tabcaption{\label{tab:fcb} Our results of decay constants (in GeV) of $B_c(nS)$ and $B^*_c(nS)$ corresponding to the three parameter sets in Table \ref{tab:parameters}, the uncertainties due to losing Lorentz covariance are listed in the parenthesis.}
\centering

\begin{tabular}{c|c|c|c|c}
\hline    
\multirow{2}{*}{ state} & \multirow{2}{*}{$J^{\textmd{P}}$} & \multicolumn{3}{c}{$f^{\textmd{QM}}_{c\bar{b}}$} \\
\cline{3-5}
&& Parameter1&  Parameter2	&  Parameter3 \\
\hline
$B_c(1\,^1S_0)$ & $0^{-}$ & 0.429(30)& 0.439(30) & 0.456(32)\\
$B_c(2\,^1S_0)$ & $0^{-}$ & 0.292(12)&0.282(13)& 0.277(13) \\
$B_c(3\,^1S_0)$ & $0^{-}$ & 0.251(5)&0.237(6)& 0.230(6)\\
$B^*_c(1\,^3S_1)$ & $1^{-}$ & 0.390(44)&0.417(51)& 0.440(56)\\
$B^*_c(2\,^3S_1)$ & $1^{-}$ & 0.294(33)&0.297(35)& 0.296(37)\\
$B^*_c(3\,^3S_1)$ & $1^{-}$ & 0.262(28)&0.257(29)& 0.253(30)\\
\hline
\end{tabular}

\end{center}

\begin{multicols}{2}

Note that $B_c(nP'_1)$ and $B_c(nP_1)$ are mixing states of $B_c(n^1P_1)$ and $B_c(n^3P_1)$,
\begin{equation} 
\begin{pmatrix}
 B_c(nP'_1) \\ B_c(nP_1)
\end{pmatrix}=
\begin{pmatrix}
 \cos\theta_{nP} & \sin\theta_{nP} \\
 -\sin\theta_{nP} & \cos\theta_{nP} 
\end{pmatrix}
\begin{pmatrix}
 B_c(n^1P_1) \\ B_c(n^3P_1)
\end{pmatrix},
\end{equation}
where $\theta_{nP}$ is the mixing angle. We always choose $B_c(nP'_1)$ to be the state nearer to $B_c(n^1P_1)$, i.e. the mixing angle is in the range $0 \leq \theta_{nP} \leq 45^\circ$. Our results of the mixing angles are listed from the second to fourth column in Table \ref{tab:mixingAngle}, and the previous quark model results using a constant strong coupling \cite{Li2019} are listed in the fifth column. We see that the mixing angle depends on the parameters. While a running coupling affects $\theta_{1P}$ very little, the mixing angles of the radial excited mesons from a running coupling are much smaller than that from a constant $\alpha_s$, i.e. the mixing of the radial excited mesons is much weaker.

As explained in section \ref{sec:decayconstant}, we get two different expressions for $f_P$ and three for $f_V$, and they depend on the momentum of the meson, due to losing Lorentz covariance. This is illustrated in Fig. \ref{fig:fp}, where the left panel is $f_{B_c(1^1S_0)}$ and the right panel is $f_{B_c(1^3S_1)}$. The dependence on the meson momentum is weak upto 2 GeV, thus the main uncertainty comes from the different expressions (Eqs. (\ref{eq:fptemporal}) and (\ref{eq:fpspacial}) for $f_P$, Eqs. (\ref{eq:fvtemporal}), (\ref{eq:fvspacialL}) and (\ref{eq:fvspacialT}) for $f_V$). We treat the central value as the predicted decay constant, and the deviation from central value as the uncertainty due to losing Lorentz covariance. Our results of decay constants of $B_c(nS)$ and $B^*_c(nS)$ corresponding to the three parameter sets and its uncertainties are listed in Table \ref{tab:fcb}. We see that the uncertainty due to losing Lorentz covariance is smaller for higher $n$ states. Comparing the results from different parameters, the uncertainty due to varying the parameter is smaller than the former one in most cases. 

Our final prediction for the decay constant together with both uncertainties are listed in Table \ref{tab:fcb-final}. We also compare our result with others. $f^{\textmd{DSE}}_{c\bar{b}}$ is the result from Dyson-Schwinger equation approach \cite{Chen2019,Chen2021}. $f^{\textmd{lQCD}}_{c\bar{b}}$ is one of the lattice QCD result \cite{Colquhoun2015}, other lattice QCD results are almost consistent with this one. The sixth and seventh columns are results from other potential models \cite{Kiselev2001,Soni2018}. The eighth column is the result from a light-front quark model \cite{Choi2015}. These results are almost consistent except that our predictions for the radial excited mesons are smaller than that of Ref. \cite{Soni2018}. The main difference is that Ref. \cite{Soni2018} uses the nonrelativistic limit van Royen and Weisskopf formula to calculate the decay constants, and this results in a larger decay constant \cite{Lakhina2006}. The reliability of our results could also be supported by the mass spectra and decay constants of the charmonium and bottomium, which are presented  in the appendixes. We can see from Table \ref{tab:mcc}, Table \ref{tab:fcc}, Table \ref{tab:mbb} and Table \ref{tab:fbb} that our results are overall consistent with other results.

\end{multicols}

\begin{center}
\tabcaption{\label{tab:fcb-final} Decay constants of $B_c(nS)$ and $B^*_c(nS)$ (in GeV). $f^{\textmd{QM}}_{c\bar{b}}$ is our prediction, where the first uncertainty is due to losing Lorentz covariance and the second uncertainty is due to varying the parematers. $f^{\textmd{DSE}}_{c\bar{b}}$ is the results from Dyson-Schwinger equation approach, $f_{B_c(1\,^1S_0)}$ and $f_{B^*_c(1\,^3S_1)}$ are from Ref. \cite{Chen2019}, $f_{B_c(2\,^1S_0)}$ and $f_{B^*_c(2\,^3S_1)}$ are from Ref. \cite{Chen2021}. $f^{\textmd{lQCD}}_{c\bar{b}}$ is the lattice QCD results \cite{Colquhoun2015}. The sixth and seventh columns are results from other potential models \cite{Kiselev2001,Soni2018}. The eighth column is the result from a light-front quark model \cite{Choi2015}.}
\centering

\begin{tabular}{c|c|c|c|c|c|c|c}
\hline    
 state & $J^{\textmd{P}}$ & $f^{\textmd{QM}}_{c\bar{b}}$  &	$f^{\textmd{DSE}}_{c\bar{b}}$\textsuperscript{\cite{Chen2019,Chen2021}}	& $f^{\textmd{lQCD}}_{c\bar{b}}$ \textsuperscript{\cite{Colquhoun2015}} & $|f|$ \hspace*{-0.3em}\textsuperscript{\cite{Kiselev2001}} & $|f|$\textsuperscript{\cite{Soni2018}} & $|f|$\textsuperscript{\cite{Choi2015}} \\
\hline
$B_c(1\,^1S_0)$ & $0^{-}$ & 0.439(30)(17) & 0.441(1) & 0.434(15) & $0.400(45)$ & 0.433 & $ 0.389^{+16}_{-3}$ \\
$B_c(2\,^1S_0)$ & $0^{-}$ & 0.282(13)(10) & 0.246(7) & -- &$0.280(50)$  & 0.356& -- \\
$B_c(3\,^1S_0)$ & $0^{-}$ &0.237(6)(14) & -- & -- & -- & 0.326 & --\\
$B^*_c(1\,^3S_1)$ & $1^{-}$ & 0.417(51)(27)& 0.431(7) & 0.422(13) & -- & 0.435 & $ 0.391^{+4}_{-5}$ \\
$B^*_c(2\,^3S_1)$ & $1^{-}$ & 0.297(35)(3)& 0.305(13) & -- & --  & 0.356& -- \\
$B^*_c(3\,^3S_1)$ & $1^{-}$ & 0.257(29)(5)&  -- & -- & -- & 0.326 & --\\
\hline
\end{tabular}

\end{center}

\begin{multicols}{2}

\section{Summary and conclusion}\label{sec:conclusion}
\vspace{0.1cm}

In summary, we calculated the decay constants of $B_c(nS)$ and $B^*_c(nS)$ mesons ($n = 1,2,3$) in the nonrelativistic quark model. Our approach can be distinguished from other quark model studies by three points:
\begin{enumerate}[itemindent=0em,label=(\arabic*)]
 \item The effect of a running strong coupling is taken into account. We use the form, Eq. (\ref{eq:runningAlphas}), which approaches the one loop running form of QCD at large $Q^2$ and saturates at low $Q^2$. A running coupling affects the wave function of Eq. (\ref{eq:SchrodingerEq}), so it has a considerable effect on the mixing angles and the decay constants.
 \item The ambiguity due to losing Lorentz covariance is discussed in detail. We get two different expressions for $f_P$ and three different expressions for $f_V$ in nonrelativistic quark model as a result of losing Lorentz covariance. The central value is treated as the prediction, and the deviation is treated as the uncertainty. We also find that the uncertainties due to losing Lorentz covariance decrease as n increases.
 \item We use three parameter sets, and the uncertainties due to varying the parameters are given. In most cases, this uncertainty is smaller than the former one.
\end{enumerate}

Comparing our results with those from other approaches, we see that they are in good agreement. In the appendixes, we compare the decay constants of charmonium and bottomium from our calculation and those from other approaches. The overall agreement also raises the credibility of our approach. All in all, the decay constants of $B_c(nS)$ and $B^*_c(nS)$ mesons ($n = 1,2,3$) are predicted, with the uncertainties well determined. And thus establish a good basis to study the decays of $B_c$ mesons.

\vspace{1em}
\acknowledgments{Acknowledgments: Thank Professor Xianhui Zhong for careful reading of the manuscript and for his useful suggestions.}

\end{multicols}
\begin{multicols}{2}

\vspace{2mm}
\centerline{\rule{80mm}{0.1pt}}
\vspace{2mm}

%
%
%
%
%

\bibliography{fBc}

\end{multicols}

\section*{Appendix A}\label{sec:appendixA}

\setcounter{equation}{0}
\renewcommand{\theequation}{A\arabic{equation}}

In this appendix, we list our nonrelativistic quark model results of the mass spectrum of charmonium in Table \ref{tab:mcc} and the decay constants of $\eta_c(nS)$ and $J/\psi(nS)$ (n=1, 2, 3) in Table \ref{tab:fcc}.
The experiment values of the vector meson decay constants ($f_V$) in Table \ref{tab:fcc} and Table \ref{tab:fbb} are estimated by
\begin{equation}\label{eq:fv-expt}
 \Gamma_{V\to e^+e^-} = \frac{4\pi \alpha^2 Q^2 *f_V^2}{3M_V},
\end{equation}
where $\Gamma_{V\to e^+e^-}$ is the decay width of the vector meson to $e^+e^-$, $\alpha$ is the fine structure constant, $Q$ is the electric charge of the constituent quark, $M_V$ is the mass of the vector meson.

\appendix
\setcounter{figure}{0}
\setcounter{table}{0}
\renewcommand{\thefigure}{A\arabic{figure}}
\renewcommand{\thetable}{A\arabic{table}}

\begin{center}
 \tabcaption{\label{tab:mcc} Mass spectrum of charmonium (in GeV). $M_{c\bar{c}}^{\text{QM}}$ is our nonrelativistic quark model result, with the parameters $m_c = 1.591 \text{ GeV}$, $\alpha_0 = 1.082$, $N_f = 4$, $\Lambda_{\text{QCD}} = 0.30 \text{ GeV}$, $b = 0.1320 \text{ GeV}^2$, $\sigma = 1.30 \text{ GeV}$, $r_c = 0.375 \text{ fm}$. Note that $N_f$ and $\Lambda_{\text{QCD}}$ are chosen according to QCD estimatation, the other parameters are tuned to fit the masses of $\eta_c(1S)$, $\eta_c(2S)$, $J/\psi(1S)$ and $\chi_{c0}(1P)$, i.e. these four masses are inputs of our model, and all the other masses are outputs. $M^{\textmd{expt.}}_{c\bar{c}}$ is the experiment value \cite{Zyla2020}. }
\begin{tabular}{c|c|c|l|l}
\hline    
$n^{2S+1}L_J$&state &$J^{\textmd{PC}}$& $M_{c\bar{c}}^{\text{QM}}$	& $M^{\textmd{expt.}}_{c\bar{c}}$ \textsuperscript{\cite{Zyla2020}}\\
\hline
$1^1S_0$& $\eta_c(1S)$& $0^{-+}$  &  2.984 (input) &  2.984(0.4) \\
$2^1S_0$& $\eta_c(2S)$& $0^{-+}$  & 3.639 (input)&  3.638(1) \\
$3^1S_0$& $\eta_c(3S)$& $0^{-+}$  & 4.054 &  \phantom{0.}-- \\
$1^3S_1$& $J/\psi(1S)$& $1^{--}$  &  3.097 (input) &  3.097(0) \\
$2^3S_1$& $\psi(2S)$& $1^{--}$ &    3.687 & 3.686(0.1) \\
$3^3S_1$& $\psi(4040)$& $1^{--}$ &  4.088  &  4.039(1) \\
$1^3P_0$& $\chi_{c0}(1P)$& $0^{++}$   & 3.415 (input)&  3.415(0.3) \\
$2^3P_0$& $\chi_{c0}(2P)$& $0^{++}$  & 3.897 &  \phantom{0.}-- \\
$3^3P_0$& $\chi_{c0}(3P)$& $0^{++}$  & 4.260 &  \phantom{0.}-- \\
$1^1P_1$& $h_{c}(1P)$& $1^{+-}$  & 3.498 &  3.525(0.1) \\
$2^1P_1$& $h_{c}(2P)$& $1^{+-}$  & 3.931 &  \phantom{0.}-- \\
$3^1P_1$& $h_{c}(3P)$& $1^{+-}$  & 4.279 &  \phantom{0.}-- \\
$1^3P_1$& $\chi_{c1}(1P)$& $1^{++}$  & 3.492 &  3.511(0.1) \\
$2^3P_1$& $\chi_{c1}(2P)$& $1^{++}$  & 3.934 &  \phantom{0.}-- \\
$3^3P_1$& $\chi_{c1}(3P)$& $1^{++}$  & 4.285 &  \phantom{0.}-- \\
$1^3P_2$& $\chi_{c2}(1P)$& $2^{++}$  &  3.534 &  3.556(0.1) \\
$2^3P_2$& $\chi_{c2}(3930)$& $2^{++}$  & 3.956 &  3.923(1) \\
$3^3P_2$& $\chi_{c2}(3P)$& $2^{++}$  & 4.299 &  -- \\
\hline
\end{tabular}

\end{center}

\begin{center}
\tabcaption{\label{tab:fcc} Decay constants of $\eta_c(nS)$ and $J/\psi(nS)$ (in GeV). $f^{\textmd{QM}}_{c\bar{c}}$ is our nonrelativistic quark model result, with the parameters listed in the caption of Table \ref{tab:mcc}. The uncertainties due to losing Lorentz covariance are listed in the parenthesis. $f^{\textmd{DSE}}_{c\bar{c}}$ is the results from Dyson-Schwinger equation (DSE) approach, where $f_{\eta_c(1\,^1S_0)}$ and $f_{J/\psi(1\,^3S_1)}$ are from Ref. \cite{Chen2019}, $f_{\eta_c(2\,^1S_0)}$ and $f_{\psi(2\,^3S_1)}$ are from Ref. \cite{Chen2021}, and the underlined values are inputs. $f^{\textmd{lQCD}}_{c\bar{b}}$ is the lattice QCD results, where $f_{\eta_c(1\,^1S_0)}$ is from Ref. \cite{McNeile2012}, and $f_{J/\psi(1\,^3S_1)}$ is from Ref. \cite{Donald2012}. The seventh and eighth columns are other potential model results \cite{Kiselev2001,Soni2018}. The ninth column is a light front quark model result \cite{Choi2015}. $f^{\textmd{SR}}_{c\bar{c}}$ is the result from QCD sum rule \cite{Becirevic2014}. $f^{\textmd{expt.}}_{c\bar{c}}$ is the experiment value and the vector meson decay constant is estimated by Eq. (\ref{eq:fv-expt}). }
\centering

\renewcommand\arraystretch{1.2}
\begin{tabular}{p{3.5em}<{\centering}|p{2.6em}<{\centering}|p{1.6em}<{\centering}|p{4em}<{\centering}|p{4em}<{\centering}|p{4em}<{\centering}|p{3.8em}<{\centering}|p{2.6em}<{\centering}|p{3.5em}<{\centering}|p{3.5em}<{\centering}|p{4.0em}<{\centering}}
\hline    
$n^{2S+1}L_J$ & state & {$J^{\textmd{PC}}$}   &	$f^{\textmd{QM}}_{c\bar{c}}$ &	$f^{\textmd{DSE}}_{c\bar{c}}$\textsuperscript{\cite{Chen2019,Chen2021}}	& $f^{\textmd{lQCD}}_{c\bar{c}}$ \textsuperscript{\cite{McNeile2012,Donald2012}} & $|f|$ \hspace*{-0.3em}\textsuperscript{\cite{Kiselev2001}} & $|f|$\textsuperscript{\cite{Soni2018}} & $|f|$\textsuperscript{\cite{Choi2015}}& $f^{\textmd{SR}}_{c\bar{c}}$\textsuperscript{\cite{Becirevic2014}}& $f^{\textmd{expt.}}_{c\bar{c}}$\textsuperscript{\cite{Zyla2020}}\\
\hline
$1\,^1S_0$ & $\eta_c(1S)$ & $0^{-+}$ & 0.447(32)& \underline{0.393} & 0.393(4) & -- & 0.350 & $ 0.353^{+22}_{-17}$ & 0.309(39) &\\
$2\,^1S_0$ & $\eta_c(2S)$ & $0^{-+}$ & 0.268(2) & 0.223(11) & -- & -- & 0.278 & -- & -- &\\
$3\,^1S_0$ & $\eta_c(3S)$ & $0^{-+}$ & 0.220(11) & -- & -- & -- & 0.249 & -- & -- &\\
$1\,^3S_1$ & $J/\psi$ & $1^{--}$ & 0.403(57) & 0.430(1) & 0.405(6) & 0.400(35) & 0.326 & $ 0.361^{+7}_{-6}$ & 0.401(46) & 0.416(8)\\
$2\,^3S_1$& $\psi(2S)$ & $1^{--}$ & 0.295(35) & \underline{0.294}(7) & -- & 0.297(26) & 0.257 & -- & -- & 0.294(5)\\
$3\,^3S_1$& $\psi(3S)$ & $1^{--}$ & 0.257(26)& -- & -- & 0.226(20) & 0.230 & -- & -- & 0.187(15)\\
\hline
\end{tabular}

\end{center}

\section*{Appendix B}  \label{sec:appendixB}

In this appendix, we list our nonrelativistic quark model results of the mass spectrum of charmonium in Table \ref{tab:mbb} and the decay constants of $\eta_b(nS)$ and $\varUpsilon(nS)$ (n=1, 2, 3) in Table \ref{tab:fbb}.

\appendix
\setcounter{figure}{0}
\setcounter{table}{0}
\renewcommand{\thefigure}{B\arabic{figure}}
\renewcommand{\thetable}{B\arabic{table}}

\begin{center}
 \tabcaption{\label{tab:mbb} Mass spectrum of bottomium (in GeV). $M_{b\bar{b}}^{\text{QM}}$ is our nonrelativistic quark model result, with the parameters $m_b = 4.997 \text{ GeV}$, $\alpha_0 = 0.920$, $N_f = 5$, $\Lambda_{\text{QCD}} = 0.30 \text{ GeV}$, $b = 0.1110\text{ GeV}^2$, $\sigma = 2.35 \text{ GeV}$, $r_c = 0.195 \text{ fm}$. Note that $N_f$ and $\Lambda_{\text{QCD}}$ are chosen by QCD estimatation, the other parameters are tuned to fit the masses of $\eta_b(1S)$, $\varUpsilon(1S)$, $\varUpsilon(2S)$ and $\chi_{b0}(1P)$, i.e. these four masses are inputs of our model, and all the other masses are outputs. $M^{\textmd{expt.}}_{b\bar{b}}$ is the experiment value \cite{Zyla2020}.}
\begin{tabular}{c|c|c|l|l}
\hline
$n^{2S+1}L_J$&state&$J^{\text{PC}}$ &	$M^{\text{QM}}_{b\bar{b}}$	& $M^{\text{expt.}}_{b\bar{b}}$ \textsuperscript{\cite{Zyla2020}}\\
\hline
$1^1S_0$ & $\eta_b(1S)$& $0^{-+}$  & \phantom{0}9.400 (input) &  \phantom{0}9.399(2) \\
$2^1S_0$ & $\eta_b(2S)$& $0^{-+}$  & 10.004 &  \phantom{0}9.999(4) \\
$3^1S_0$ & $\eta_b(3S)$& $0^{-+}$  & 10.324 &  \phantom{00.}-- \\
$1^3S_1$ & $\varUpsilon(1S)$& $1^{--}$  &  \phantom{0}9.460 (input) & \phantom{0}9.460(0.3) \\
$2^3S_1$ & $\varUpsilon(2S)$& $1^{--}$  & 10.023 (input) & 10.023(0.3) \\
$3^3S_1$ & $\varUpsilon(3S)$& $1^{--}$  & 10.336 & 10.355(1) \\
$4^3S_1$ & $\varUpsilon(4S)$& $1^{--}$  & 10.573 & 10.579(1) \\
$1^3P_0$ & $\chi_{b0}(1P)$& $0^{++}$  & \phantom{0}9.859 (input) & \phantom{0}9.859(1) \\
$2^3P_0$ & $\chi_{b0}(2P)$& $0^{++}$  & 10.224 & 10.233(1) \\
$3^3P_0$ & $\chi_{b0}(3P)$& $0^{++}$  & 10.481 & \phantom{00.}-- \\
$1^1P_1$ & $h_{b}(1P)$ & $1^{+-}$  &  \phantom{0}9.903 &  \phantom{0}9.899(1) \\
$2^1P_1$ & $h_{b}(2P)$ & $1^{+-}$  & 10.244 &  10.260(1) \\
$3^1P_1$ & $h_{b}(3P)$ & $1^{+-}$  & 10.493 &  \phantom{00.}-- \\
$1^3P_1$ & $\chi_{b1}(1P)$& $1^{++}$  & \phantom{0}9.896 & \phantom{0}9.893(1) \\
$2^3P_1$ & $\chi_{b1}(2P)$& $1^{++}$  & 10.242 & 10.255(1) \\
$3^3P_1$ & $\chi_{b1}(3P)$& $1^{++}$  & 10.493 & 10.513(1) \\
$1^3P_2$ & $\chi_{b2}(1P)$& $2^{++}$  & \phantom{0}9.921 & \phantom{0}9.912(1) \\
$2^3P_2$ & $\chi_{b2}(2P)$& $2^{++}$  & 10.255 & 10.269(1) \\
$3^3P_2$ & $\chi_{b2}(3P)$& $2^{++}$  & 10.502 & 10.524(1) \\
$1^3D_2$ & $\varUpsilon_2(1D)$& $2^{--}$ &  10.152 & 10.164(1) \\
\hline
\end{tabular}

\end{center}

\begin{center}
\tabcaption{\label{tab:fbb} Decay constants of $\eta_b(nS)$ and $\varUpsilon(nS)$ (in GeV). $f^{\textmd{QM}}_{b\bar{b}}$ is our nonrelativistic quark model result, with the parameters listed in the caption of Table \ref{tab:mbb}. The uncertainties due to losing Lorentz covariance are listed in the parenthesis. $f^{\textmd{DSE}}_{b\bar{b}}$ is the results from Dyson-Schwinger equation (DSE) approach, where $f_{\eta_b(1\,^1S_0)}$ and $f_{\varUpsilon(1\,^3S_1)}$ are from Ref. \cite{Chen2019}, $f_{\eta_b(2\,^1S_0)}$ and $f_{\varUpsilon(2\,^3S_1)}$ are from Ref. \cite{Chen2021}, and the underlined values are inputs. $f^{\textmd{lQCD}}_{c\bar{b}}$ is the lattice QCD results, where $f_{\eta_b(1\,^1S_0)}$ is from Ref. \cite{McNeile2012}, $f_{\varUpsilon(1\,^3S_1)}$ and $f_{\varUpsilon(2\,^3S_1)}$ are from Ref. \cite{Colquhoun2015a}.  The seventh and eighth columns are other potential model results \cite{Kiselev2001,Soni2018}. The ninth column is a light front quark model result \cite{Choi2015}. $f^{\textmd{expt.}}_{b\bar{b}}$ is the experiment value and the vector meson decay constant is estimated by Eq. (\ref{eq:fv-expt}).}
\centering

\renewcommand\arraystretch{1.2}
\begin{tabular}{p{3.5em}<{\centering}|p{2.6em}<{\centering}|p{1.6em}<{\centering}|p{4em}<{\centering}|p{4em}<{\centering}|p{4em}<{\centering}|p{3.8em}<{\centering}|p{2.6em}<{\centering}|p{3.5em}<{\centering}|p{4.0em}<{\centering}}
\hline
$n^{2S+1}L_J$ & state & {$J^{\textmd{PC}}$} &	$f^{\textmd{QM}}_{b\bar{b}}$    &	$f^{\textmd{DSE}}_{b\bar{b}}$\textsuperscript{\cite{Chen2019,Chen2021}}	& $f^{\textmd{lQCD}}_{b\bar{b}}$ \textsuperscript{\cite{McNeile2012,Colquhoun2015a}} & $|f|$ \hspace*{-0.3em}\textsuperscript{\cite{Kiselev2001}} & $|f|$\textsuperscript{\cite{Soni2018}}& $|f|$ \hspace*{-0.3em}\textsuperscript{\cite{Choi2015}}& $f^{\textmd{expt.}}_{b\bar{b}}$ \textsuperscript{\cite{Zyla2020}}\\
\hline
$1\,^1S_0$ & $\eta_b(1S)$ & $0^{-+}$ & 0.749(41)& \underline{0.667} & 0.667(6) & -- & 0.646 & $0.605^{+32}_{-17}$ &\\
$2\,^1S_0$ &$\eta_b(2S)$ & $0^{-+}$ & 0.441(14)& 0.488(8) & -- & -- & 0.519 & -- & \\
$3\,^1S_0$ &$\eta_b(3S)$ & $0^{-+}$ & 0.356(7)& -- & -- & -- & 0.475 & -- & \\
$1\,^3S_1$ &$\varUpsilon(1S)$ & $1^{--}$ & 0.712(78)& 0.625(4) & 0.649(31) & 0.685(30) & 0.647 & $0.611^{+6}_{-11}$ & 0.715(10)\\
$2\,^3S_1$ &$\varUpsilon(2S)$ & $1^{--}$ & 0.460(48)& \underline{0.498}(6) & 0.481(39) & 0.469(21) & 0.519 & -- & 0.497(9)\\
$3\,^3S_1$ &$\varUpsilon(3S)$ & $1^{--}$ & 0.381(38)& -- & -- & 0.399(17) & 0.475 & -- & 0.425(8)\\
\hline
\end{tabular}

\end{center}

\end{document}